\documentclass[twocolumn]{revtex4}

\usepackage{graphicx}
\usepackage{amsmath,amssymb}
\usepackage{color}

\begin{document}

\def\apjl{\ {Astrophys. J.}\ } \def\apjs{\ {Astrophys. J. Suppl.}\ }


\title{Energetics of high-energy cosmic radiations}

\author{Kohta Murase} \affiliation{Department of Physics; Department
  of Astronomy \& Astrophysics; Center for Particle and Gravitational
  Astrophysics, Pennsylvania State University, University Park,
  Pennsylvania 16802, USA} \affiliation{Yukawa Institute for
  Theoretical Physics, Kyoto University, Kyoto, Kyoto 606-8502, Japan}
\author{Masataka Fukugita} \affiliation{Institute for Advanced Study,
  Princeton, New Jersey 08540, USA} \affiliation{Kavli Institute for
  the Physics and Mathematics of the Universe, University of Tokyo,
  Kashiwa 277 8583, Japan}


\begin{abstract}
The luminosity densities of high-energy cosmic radiations are studied
to find connections among the various components, including
high-energy neutrinos measured with IceCube and gamma rays with the
{\it Fermi} satellite.  Matching the cosmic-ray energy generation rate
density in a GeV-TeV range estimated for Milky Way with the
ultrahigh-energy component requires a power-law index of the spectrum,
$s_{\rm cr}\approx2.1-2.2$, somewhat harder than $s_{\rm cr}\approx2.3-2.4$ 
for the local index derived from the AMS-02 experiment. 
The soft GeV-TeV cosmic-ray spectrum extrapolated to higher energies can be compatible
with PeV cosmic rays inferred from neutrino measurements, but
overshoots the CR luminosity density to explain GeV-TeV gamma rays.
The extrapolation from ultrahigh energies with a hard spectrum, on the
other hand, can be consistent with both neutrinos and gamma-rays.  
These point towards either reacceleration of galactic cosmic rays or the presence of
extragalactic sources with a hard spectrum. 
We discuss possible cosmic-ray sources that can be added.
\end{abstract}

\pacs{95.85.Ry, 98.70.Sa, 98.70.Vc\vspace{-0.3cm}}
\maketitle

%
\section{Introduction}

Cosmic rays (CRs) carry the energy about 1 eV per cubic centimetre in
the solar neighbourhood.  This amounts to $\Omega_{\rm cr}\sim10^{-11.4}$ 
when the local energy density is extended to the
entire Milky Way galaxy and integrated over the optical luminosity
function of galaxies, assuming that the CR energy is proportional to
optical luminosity of galaxies~\cite{Fukugita:2004ee}.  It is known,
however, that CRs leak from our Galaxy in a time scale about
$\sim10-100$~Myr.  When this is taken into account, assuming that
leaked CRs survive without significant energy losses, the global CR
energy density amounts to $\Omega_{\rm cr}\sim10^{-8.3}$~\cite{Fukugita:2004ee}.  
This energy that represents the CR generation is about 20\% of the cumulative amount of
kinetic energies produced in core-collapse supernovae (ccSNe)
integrated to higher redshifts, $\Omega_{\rm sn, ke}\sim10^{-7.3}$~\cite{Fukugita:2004ee}.  
Such an energetics consideration endorses that the production of CRs is associated with
the star-formation activity in galaxies.

In the latest years much information relevant to extragalactic CRs
becomes available at very high energies. This would raise the question
as to connections among various extragalactic cosmic particles and CRs
that are locally observed~\cite{Murase:2013rfa,Katz:2013ooa}.  The
observations of high-energy CRs are supplemented by unprecedentedly
accurate knowledge of local CRs in the GeV-TeV region by
PAMELA~\cite{Adriani:2011cu} and by AMS-02~\cite{Aguilar:2015ooa},
which leads us to infer accurately the propagation of Galactic CRs,
and thus would in turn make clear the position of Galactic and
extragalactic CRs in the universe.

Ultrahigh-energy (UHE) CRs in excess of $10^{18.5}$~eV are likely to
be extragalactic: further to the fact that they cannot be confined in
the Galaxy, their spectrum shows a sharp decrease above the energy
$5\times 10^{19}$eV, as observed in the Pierre Auger Observatory
\cite{Aab:2017njo} and the Telescope Array
\cite{TheTelescopeArray:2018dje}, which can be ascribed to the
Greisen-Zatsepin-Kuzmin (GZK) and photodisintegration cutoffs, and
indicates the origin of UHE CRs at a great distance of several tens of Mpc.

In addition, observations have been made for GeV-TeV gamma rays with
the {\it Fermi} telescope (e.g.,~\cite{Ackermann:2014usa}), and for
TeV-PeV cosmic neutrinos with the IceCube experiment
(e.g.,~\cite{Aartsen:2017mau}).  These experiments tell us about the
high-energy CRs as primaries to be compared with the direct CR
observation at lower energies.

In this paper we focus on luminosity densities of GeV-TeV and UHE CRs,
and those needed to account for high-energy gamma-ray and neutrino
observations.  We study physical connections among these components.
This consideration hints us to clarify the origin of high-energy
cosmic particles. 
We consider that a typical error of our argument is no greater than 0.3 dex, 
unless otherwise explicitly noted, from a number of cross checks we have performed.

We take $H_0=70~{\rm km}~{\rm s}^{-1}~{\rm Mpc}^{-1}$, $\Omega_m=0.3$
and $\Omega_\Lambda=0.7$ for cosmological parameters.

\section{Energetics of Cosmic Radiations}
\subsection{Galactic cosmic rays}
\subsubsection{Galactic cosmic-ray luminosity}
The CR proton flux derived by AMS-02 is given approximately
as~\cite{Aguilar:2015ooa}
\begin{equation}
E^2\Phi_{\rm cr}=1.8\times{10}^{-2}~{\rm GeV}~{\rm cm}^{-2}~{\rm s}^{-1}
~{\rm sr}^{-1}~{\left(\frac{E}{300~{\rm GeV}}\right)}^{2-\gamma_{\rm cr}},
\end{equation}
for the kinetic energy $E\gtrsim50$~GeV, where $\gamma_{\rm cr}$ is
the spectral index of CRs observed on Earth: $\gamma_{\rm cr}\approx2.85$ 
below $\sim300$~GeV hardens to $\gamma_{\rm cr}\approx2.72$ above 
$\sim300$~GeV, This hardening has been known
from the CREAM~\cite{Yoon:2011aa} and PAMELA
experiments~\cite{Adriani:2011cu}.  The helium flux is known to be
harder by $\Delta\gamma_{\rm cr}=-0.08$, corresponding to 
$\gamma_{\rm  cr}\approx2.78$ below $\sim250$~GV that hardens to 
$\gamma_{\rm  cr}\approx2.66$ above $\sim250$~GV~\cite{Aguilar:2015ctt}.
We use Refs.~\cite{Aguilar:2015ooa,Aguilar:2015ctt} for proton and
helium fluxes above $46$~GV.  For lower-energies, where the solar
modulation is more important, we adopt Ref.~\cite{Boschini:2017fxq}
that considers the latest Voyager data~\cite{2013Sci...341..150S}.
See also Ref.~\cite{Corti:2015bqi} for a detailed study on solar
modulation effects.

The Galactic CR energy density is written
\begin{equation}
EdU_{\rm cr}/dE=dU_{\rm cr}/d\ln E =4\pi E^2\Phi_{\rm cr}/v,
\label{eq:CRenergy}
\end{equation}
where $v$ is the particle velocity.  The total energy density of CR
protons and helium nuclei is estimated to be $U_{\rm cr}=\int dE \,
(dU_{\rm cr}/dE)\approx1~{\rm eV}~{\rm cm}^{-3}$, consistent with the
value based on Ref.~\cite{DiMauro:2015jxa}, taking account of the
solar modulation using the Fisk model with the solar potential 700~MV.

To evaluate the high-energy CR luminosity of the Milky Way, we take
account of the fact that CRs eventually escape from the Galactic disc
and the larger CR halo region.  
We here introduce the grammage along the CR path length 
$X_{\rm esc}\equiv\int dl\,n(l)\mu$, where $n(l)$ is the nucleon density and $\mu\approx1.4m_p$
is the mean mass of gas.  In the steady state, $X_{\rm esc}$ is
related to the CR residence time $t_{\rm esc}$, as 
$X_{\rm esc}(R)=\bar{n}\mu vt_{\rm esc}(R)$, where $\bar{n}$ 
is the mean nucleon density in the CR confinement volume in the Milky Way.  
The CR halo, which is typically $h\sim1-10$~kpc, is larger than the 
scale hight of the Galactic disk ($\sim300$~pc).  Correspondingly, 
$\bar{n}$ should be lower than the averaged density in 
the Galactic disc, while the exact value is uncertain as it 
depends on the CR halo size~\cite{Fukugita:2004ee} that is uncertain too.
Instead, the grammage that consists of the product of $\bar{n}$ 
and $t_{\rm esc}$ can be determined better by CR data~\cite{Blum:2013zsa}.  From
the ratio of boron to carbon fluxes~\cite{Adriani:2014xoa,Aguilar:2016vqr} the 
grammage (e.g.,~\cite{Blum:2013zsa,Giacinti:2015hva}) traversed by CRs is estimated to be,
\begin{equation}
X_{\rm esc}(R)\approx8.7~{\rm g}~{\rm cm}^{-2}~{\left(\frac{R}{10~{\rm GV}}\right)}^{-\delta},
\label{eq:grammage}
\end{equation}
where $R=cp/Ze$ is the rigidity.  Whereas a single power law with
$\delta=0.4$ gives a reasonable fit (with 20\% accuracy) to the
grammage deduced from the boron-to-carbon ratio data for
$R>5$~GV~\cite{Blum:2013zsa}, the recent AMS-02 measurement indicates
a lower value, $\delta=0.333\pm0.014({\rm fit})\pm0.005(\rm syst)$
above 65~GV~\cite{Aguilar:2016vqr}.  Noting that $\delta$ appears
decreasing as energy increases~\cite{Aguilar:2018njt}, we take a
broken power law with $\delta=0.46$ for $R<250~{\rm GV}$ and
$\delta=0.33$ for $R\geq250~{\rm GV}$.  The hardening of the CR
spectrum is then translated to this flatter energy dependence of the
grammage at a higher rigidity: with the index of the CR injection
spectrum $s_{\rm cr}=\gamma_{\rm cr}-\delta=2.39$ (for the proton) and
$2.33$ (for the helium), $\delta$ can give proper spectral indices of
CRs for a wide range of the energy (see also
Ref.~\cite{Aguilar:2018njt}).  Our conclusions are unchanged by the
choice of $\delta$ within the uncertainty.  We remark that for
$R\lesssim20$ GV the {\it Fermi} measurement of gamma rays from nearby
molecular clouds gives different indices, e.g., $\gamma_{\rm cr}\approx2.9$ 
for $R\gtrsim10-20$~GV~\cite{Neronov:2017lqd}, implying a steeper high-energy CR spectrum.

Now let us estimate the CR luminosity of the Milky Way.  In the 
steady state, the differential CR luminosity satisfies 
$E(dL_{\rm cr}/dE)t_{\rm esc} =E(dU_{\rm cr}/dE)V_{\rm halo}$, where $V_{\rm halo}$ is the halo volume.  This can be
rewritten $E(dL_{\rm cr}/dE)X_{\rm esc} =E(dU_{\rm cr}/dE)M_{\rm gas}$, 
where $M_{\rm gas}$ is the total gas mass contained in the CR halo.  
In general, the gas mass consists of 
$M_{\rm gas}=M_{\rm cold}+M_{\rm warm/hot}$, where $M_{\rm cold}$ 
is the cold gas mass in the disc region.  
The so-called missing baryon problem~\cite{Fukugita:1997bi} implies that the latter, 
i.e., warm or hot circumgalactic mass, is significant in a scale of 
the virial radius of the Milky Way, which is about 250~kpc.  
Refs.~\cite{2013ApJ...762...20F,Werk:2014fza,2015ApJ...807..103Z}
suggest $M_{\rm warm/hot}\approx(1-4)\times{10}^8~M_\odot$ within
15~kpc and $M_{\rm warm/hot}\approx(2-3)\times{10}^9~M_\odot$ within
50~kpc.  Thus we can safely ignore the circumgalactic gas mass in a
scale of the CR halo with $h\sim1-10$~kpc.
The stellar mass of the Milky Way is estimated $M_*=5.1\times{10}^{10}~M_\odot$ (e.g.,
\cite{Licquia:2014rsa}).  The cold gas mass (H I, He I and molecular
gas) estimated from HIPASS and CO surveys, $\Omega_{\rm
  gas}/\Omega_*=0.00078/0.0027\simeq29$\%~\cite{Fukugita:2004ee},
which is consistent with an estimate for the gas mass of the Milky Way
within a factor of 2 (e.g.,~\cite{2016PASJ...68....5N}).  Taking these
uncertainties into account, we take the gas mass fraction to be
$15-30$\% of the stellar mass: $M_{\rm gas}\approx M_{\rm cold}
=(0.75-1.5)\times{10}^{10}~M_\odot$.  For the steady state the
Galactic CR proton luminosity per logarithmic energy range is, using
Eq.~(\ref{eq:grammage}),
\begin{eqnarray}
E\frac{dL_{\rm cr}}{d E}&=&E\frac{dU_{\rm cr}}{dE}\frac{vV_{\rm halo}}{t_{\rm esc}(E)}\nonumber\\
&=&E\frac{dU_{\rm cr}}{dE}\frac{vM_{\rm gas}}{X_{\rm esc}(E)}\nonumber\\
&\simeq&1.5\times{10}^{40}~{\rm erg}~{\rm s}^{-1}\nonumber\\
&\times&{\left(\frac{M_{\rm gas}}{10^{10}~M_\odot}\right)}
{\left(\frac{E}{10~{\rm GeV}}\right)}^{2-s_{\rm cr}}.\cr
&&
\label{cr}
\end{eqnarray}
We note that the above quantity is the differential luminosity 
multiplied by the relevant energy, which should be smaller than the integrated luminosity, 
$L_{\rm cr}=\int dE\,(dL_{\rm cr}/dE)$.  With the helium contribution, 
Eq.~(\ref{cr}) leads to $L_{\rm cr}\simeq8.5\times{10}^{40}~{\rm erg}~{\rm s}^{-1}$, consistent with the estimate in 
Ref.~\cite{2012APh....39...52D} within 20\%. 

We remark that the derived CR luminosity is based on local CRs in the
solar neighbourhood.  Observations of gamma rays from nearby molecular
clouds~\cite{Neronov:2017lqd} give a result consistent with this.
The CR density may be larger toward the Galactic centre.  If we
assume that the CR production is proportional to the radio
emission~\cite{1985A&A...153...17B}, the total density of CRs may be
larger than our estimate by a factor of 2~\cite{Fukugita:2004ee}.  
The recent observations in GeV-TeV gamma rays suggest that
the CR density is higher by a factor of 5 within 100~pc of the
Galactic centre~\cite{Abramowski:2016mir}, or by a factor of $2-4$
within the 3~kpc region~\cite{Acero:2016qlg,Gaggero:2017jts}.  These
mean that the CR luminosity enhancement is at most modest, only up
to by a factor of 2 when averaged over the galaxy.

Indeed, if one invokes a propagation model such as GALPROP, in
which an inhomogeneous distribution of CRs including the
enhancement around the Galactic centre, can be captured~\cite{Acero:2016qlg,Gaggero:2017jts}, 
one obtains $E(dL_{\rm cr}/dE)\simeq1.3\times{10}^{40}~{\rm erg}~{\rm s}^{-1}$
around 10~GeV~\cite{2010ApJ...722L..58S}.  Our result in
Eq.~(\ref{cr}), without taking account of the enhancement by the CR inhomogeneity,
agrees with variants of estimates to within a factor of 2.
This is an example that our estimate of the uncertainty 
in the CR luminosity, typically less than 0.3 dex, in fact, holds.

\subsubsection{Consistency with traditional estimates}
Let us note here the consistency between CR and acceleration in
remnants of ccSNe. 
If we take the conventional ccSN rate, $\rho_{\rm sn}=1/30~{\rm yr}^{-1}$~\cite{Adams:2013ana}, 
we find the average CR luminosity generated in the Milky Way galaxy,
\begin{equation}
E\frac{dL_{\rm cr}}{d E}\simeq2.1\times{10}^{40}~{\rm erg}~{\rm s}^{-1}
~\left(\epsilon_{\rm cr}{\mathcal E_{\rm sn}}\over10^{50}{\rm erg}\right)
~{\left(\frac{E}{10~{\rm GeV}}\right)}^{-0.39},
\label{cr2}
\end{equation}
taking the CR spectrum from AMS-02, the kinetic energy of a ccSN
${\mathcal E}_{\rm sn}\sim10^{51}$~erg,  
and the energy fraction carried by CRs $\epsilon_{\rm cr}\sim0.1$. 
This is consistent with Eq.~(\ref{cr}). 
Thus, ccSNe may well account for the entire CR luminosity 
with the energy fraction carried by CRs, $\epsilon_{\rm cr}\sim10$\%.

The abundance of the secondary nuclei, boron, lithium and beryllium, indicate
$t_{\rm esc}\sim15-100$~Myr~\cite{Lipari:2014zna,Blum:2010nx}.  In
particular, measurements of the beryllium ratio lead to
$t_{\rm esc}=15\pm1.6$~Myr with the aid of the leaky box model: the
diffusion model predicts a longer time scale,
$t_{\rm esc}\sim30-100$~Myr, which depends on the CR halo size,
$h$~\cite{Lipari:2014zna}. 
This is also consistent with Ref.~\cite{1998ApJ...506..335W}.
These are at least by three orders of
magnitude longer than the crossing time,
$t_{\rm cross}=h/c\simeq13~{\rm kyr}~{(h/4~{\rm kpc})}$.  This escape
time means that a dominant part of generated CRs escape.  Namely, the
fraction of CRs that reside in galaxies is $\sim t_{\rm esc}/t_H$
times the generated CRs (where $t_H=1/H_0$ is the Hubble time).  
Under the assumption that the CR halo size
is energy independent, 
the simple diffusion model
(e.g.,~\cite{Taillet:2002ub,Lipari:2014zna}) gives
\begin{equation}
t_{\rm esc}(R)\sim60~{\rm Myr}~{(c/v)}{(R/1~{\rm GV})}^{-\delta},
\label{eq:escapetime}
\end{equation} 
which is consistent with the diffusion constant at 1 GV, 
$D(1{\rm GV})\sim4\times{10}^{28}~{\rm cm}^2~{\rm s}^{-1}$ for $h=4$~kpc.  
If we assume that CRs reside in the CR halo with the volume of $V_{\rm halo}=2\pi
r_{\rm MW}^2h$ with $r_{\rm MW}=20$~kpc and $h=4$~kpc, we obtain
Galactic CR proton luminosity per logarithmic energy,
\begin{equation}
E{dU_{\rm cr}\over dE} \frac{V_{\rm halo}}{t_{\rm esc}(E)}
\sim2\times{10}^{40}~{\rm erg}~{\rm s}^{-1}~{\left(\frac{E}{10~{\rm GeV}}\right)}^{-0.39},
\label{eq:galCRlum}
\end{equation}
consistent with Eqs.~(\ref{cr}) and (\ref{cr2}). 
Eqs.~(\ref{cr}) and (\ref{eq:galCRlum}) may not be fully independent 
because $D$ and $h$ are usually determined by exploiting a propagation model 
and using information on the ratio of secondary and primary CR fluxes.
The advantage of Eq.~(\ref{cr}), over Eq.~(\ref{eq:galCRlum}), is that the CR 
luminosity is expressed with $M_{\rm gas}$ and $X_{\rm esc}$ that 
do not explicitly refer to the uncertain CR halo size, $h$.
Here it is more useful to show both derivations to see 
its consistency with the traditional expression (i.e., Eq.~\ref{eq:galCRlum}).

\begin{figure}[t]
\includegraphics[width=3.50in]{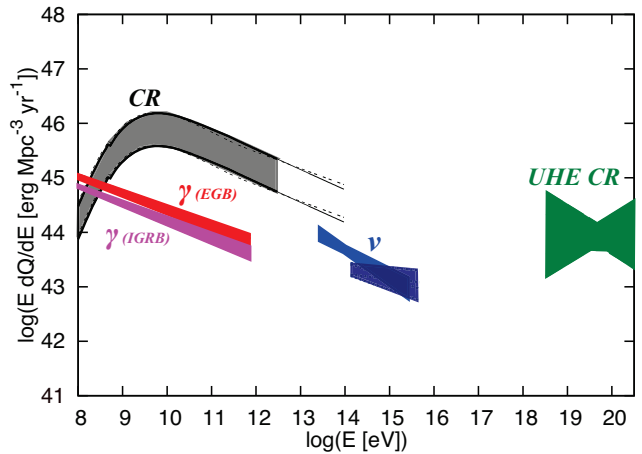}
\caption{Local ($z=0$) CR energy generation rate density estimated 
from the AMS-02, Voyager and other relevant CR data. 
UHE CR luminosity density is based on the Auger experiment.   
High-energy neutrino and gamma-ray luminosity densities use 
IceCube and {\it Fermi} Large Area Telescope, respectively. 
With the IceCube data both results from the global analysis 
and the upgoing muon neutrino analysis are included.
For gamma rays both total ``extragalactic gamma-ray
background'' (EGB: including resolved gamma-ray sources) and 
``isotropic diffuse gamma-ray background (IGRB)'' are displayed. 
The star-formation history is assumed for the neutrino and 
gamma-ray luminosity densities as discussed in the text.
\label{fig1}
}
\vspace{-1.\baselineskip}
\end{figure}

\subsubsection{Global cosmic-ray energy generation rate density}
With the CR dominantly produced in ccSNe we may assume that its energy
generation rate density is proportional to the star-formation rate $\psi$.  We
may take the star-formation rate of the Milky Way,
${\psi}_{\rm MW}\approx1.7\pm0.2~M_\odot~{\rm yr}^{-1}$~\cite{Licquia:2014rsa}, which is consistent with the
global value at $z=0$, $\psi=0.015M_\odot~{\rm Mpc}^{-3}~{\rm yr}^{-1}$~\cite{Madau:2014bja}
within a factor of 2, when scaled the Milky Way luminosity to the global luminosity density of galaxies.

Using Eq.~(\ref{cr}), the global CR energy generation rate density,
or simply CR luminosity density, $Q_{\rm cr}$ is given per logarithmic energy, as
\begin{eqnarray}
E{dQ_{\rm cr}\over dE}&=&{dQ_{\rm cr}\over d\ln E}=\frac{\psi}{\psi_{\rm MW}}{dL_{\rm cr}\over d\ln E}\nonumber\\
&\simeq&{10}^{45.9\pm0.3}~{\rm erg}~{\rm Mpc}^{-3}~{\rm yr}^{-1}~{\left(\frac{E}{10~{\rm GeV}}\right)}^{2-s_{\rm cr}},\cr
&&
\end{eqnarray}
including the helium contribution.  

The result is shown in Fig.~1 in the (grey) shaded region for
$E<3$~TeV, where the AMS-02 data are available.  A factor of 4
uncertainty is implied here to represent various uncertainties, 
such as the CR density enhancement around the Galactic centre region, 
the gas mass estimate, star-formation rate, and so on.  
The calculation with the single power-law grammage
with $\delta=0.4$ is shown with the double dashed lines, which differ
little from the broken power-law case shown with the solid lines,
showing impacts of a different power-law index.
In this figure we also depict the UHE CR from 
Auger, the gamma-ray luminosity density from the {\it Fermi}
Large Area Telescope~\cite{Ackermann:2014usa}, and the high-energy
neutrino luminosity density from IceCube~\cite{Aartsen:2017mau}, 
to compare them with the CR luminosity density, as we discuss in later subsections.  
Table I summarises the luminosity and energy densities of high-energy
cosmic particles.

Our results imply that the source spectral index for the proton
component is larger than the nominal value expected by the diffusive
shock acceleration mechanism for nonrelativistic quasiparallel shocks
(i.e., $s_{\rm cr}=2.0$). 
Our larger index is consistent with the recent result based on the leaky box model~\cite{Webber:2018bom}, 
and the observations of young supernova remnants such as Cas A~\cite{Ahnen:2017uny}.
Such a steeper spectrum might point to additional processes that may play a role in the CR acceleration~(e.g.,~\cite{1996A&A...314.1010K,Ohira:2009rd,Bell:2011cs,Ohira:2012dn}).

We remark that $s_{\rm cr}\approx2.3-2.4$ derived from the AMS-02 experiment is the local spectral index.  
The CR spectral index may vary spatially across the Milky Way, which can be probed by gamma-ray observations. 
For instance, GeV gamma-ray observations suggest harder indices in the inner regions of the Milky Way~\cite{Acero:2016qlg}.  
Somewhat a harder global index such as $s_{\rm cr}\sim2.2$ may be allowed by the anisotropic diffusion~\cite{Cerri:2017joy}, 
although indices required for the perpendicular diffusion, $\delta\sim0.5$, seem larger than the value suggested 
by the AMS-02 experiment~\cite{Aguilar:2016vqr}.  
On the other hard, a relatively soft spectrum with $s_{\rm cr}\sim2.4$ is also inferred from the gamma-ray data 
of the starburst galaxy Arp 220~\cite{Peng:2016nsx}. 
The global index is by no means definitive in the state-of-the-art experiments.
For this study we take the local value inferred from the AMS-02 experiment literally.
We note that a 0.1 difference in the index causes a 0.5~dex change in the luminosity
density at $10^{19}$~eV when extrapolated from $10^{14}$~eV. 
This is somewhat beyond what we tolerate at UHE energies.  

\begin{table}[t]
\begin{center}
\caption{Luminosity densities $Q$ in units of ${\rm erg}~{\rm Mpc}^{-3}~{\rm yr}^{-1}$ 
and energy densities $\Omega$ in units of the critical density $\varrho_{\rm crit}c^2$
of CRs, UHE CRs, high-energy neutrinos, and gamma rays.
For redshift evolution the star-formation rate history is assumed, 
as discussed in the text. 
\label{tb1}
}
\scalebox{0.95}{
\begin{tabular}{|c|c|c|c|c|c|c|}
\hline & CR & UHE CR &  $\nu$ (global) & $\nu$ (upgoing) & $\gamma$ (total)& $\gamma$ (IGRB)\\
\hline $Q$ & ${10}^{46.6}$ & ${10}^{44.5}$ & ${10}^{44.3}$ & ${10}^{43.8}$ & ${10}^{45.52}$ & ${10}^{45.31}$\\
\hline $\Omega$ & ${10}^{-8.3}$ & ${10}^{-11.5}$ & ${10}^{-10.7}$ & ${10}^{-11.2}$ & ${10}^{-9.44}$ & ${10}^{-9.64}$\\
\hline
\end{tabular}
}
\end{center}
\end{table}

%
\subsection{Extragalactic cosmic rays}
UHE CRs that cannot be confined in the Milky Way would represent
extragalactic CRs.  To estimate the luminosity density, we
assume that UHE CRs are composed dominantly of protons.  The dominance
of protons or light nuclei is suggested in the energy range
near the ankle~\cite{Aab:2017njo,TheTelescopeArray:2018dje}, and we
find that this assumption would not result in too significant an error
in our order of magnitude estimate. The UHE CR flux measured by Auger
is written~\cite{Verzi:2017hro}
\begin{eqnarray}
E^2\Phi_{\rm uhecr}&=&2.43\times{10}^{-8}~{\rm GeV}~{\rm cm}^{-2}~{\rm s}^{-1}~{\rm sr}^{-1}~{\left(\frac{E}{E_{\rm ankle}}\right)}^{2-\gamma_{\rm cr}}\nonumber\\
&\times&\left\{\begin{array}{ll} 
1
& \mbox{($E<E_{\rm ankle}$)}\\
\left[\frac{1+{(E_{\rm ankle}/E_{\rm supp})}^{\Delta\gamma_{\rm cr}}}{1+{(E/E_{\rm supp})}^{\Delta\gamma_{\rm cr}}}\right]
& \mbox{($E\geq E_{\rm ankle}$)}\\
\end{array} \right.
\label{eq:uhecr}
\end{eqnarray}
where $\gamma_{\rm cr}=3.29$ below the ankle energy
$E_{\rm ankle}=(4.82\pm0.8)\times 10^{18}$~eV, and
$\gamma_{\rm cr}=2.60$ above it; 
$E_{\rm supp}=(4.21\pm0.78)\times10^{19}$~eV is the energy of suppression 
and $\Delta\gamma_{\rm cr}=3.1$.

Using the observed UHE CR flux and their characteristic energy-loss lengths,
$ct_{\rm loss}\sim100-1000$~Mpc at $E\sim10^{19}-10^{20}$~eV, 
the UHE CR luminosity density is roughly
\begin{eqnarray}
E{dQ_{\rm uhecr}\over dE}&\approx&\frac{4\pi E^2\Phi_{\rm uhecr}}{ct_{\rm loss}}\nonumber\\ 
&\simeq&0.9\times{10}^{44}~{\rm  erg}~{\rm Mpc}^{-3}~{\rm yr}^{-1}
~{\left(\frac{ct_{\rm loss}}{1~{\rm Gpc}}\right)}^{-1}.\cr
&&
\end{eqnarray}
at around $E=10^{19}$~eV.  
Calculations of the propagation following Refs.~\cite{Murase:2008sa,Zhang:2017moz},
taking into account energy losses due to the photonuclear and Bethe-Heitler processes, give 
\begin{equation}
\footnotesize{
E{dQ_{\rm uhecr}\over d E}\approx{10}^{43.8\pm0.2}~{\rm erg}~{\rm Mpc}^{-3}~{\rm yr}^{-1}~C(s_{\rm cr}){\left(\frac{E}{10^{19.5}~{\rm eV}}\right)}^{2-s_{\rm cr}},
\label{uhebudget}
}
\end{equation}
where $C(s_{\rm cr})={\rm max}[1,s_{\rm cr}-1]$ for $1.5\lesssim
s_{\rm cr}\lesssim2.5$, being consistent with
Refs.~\cite{Katz:2008xx,Decerprit:2011qe} within 50\%.  The result on
the UHE CR luminosity density depends on the fitting range, which
reflects details of modeling of the Galactic to extragalactic
transition.  It is slightly affected by the redshift evolution of
sources, but is not very important, for UHE CRs from high redshift
sources are attenuated by energy losses.  The interpretation of 
the shower maximum in air-shower experiments is under debate, and 
the UHE CR composition is unclear due to uncertainties in hadronic 
interaction models.  
The UHE CR spectrum can be fitted not only by protons but also by medium 
heavy or heavy nuclei, and the composition also affects precise values of 
the UHECR generation rate density.
However, because energy loss lengths of different nuclei are comparable 
at $E\sim{10}^{19}$~eV (e.g.,~\cite{Allard:2011aa}), their differential UHE CR generation
rate densities are similar in this energy range. For example, at $E\sim{10}^{19}$~eV, 
energy loss lengths of nuclei lighter than nitrogen or oxygen are slightly shorter than
that of protons. Heavier nuclei have longer but still comparable loss lengths thanks to unavoidable
energy losses by the cosmic expansion, which results in somewhat smaller values of the UHE CR generation
rate density. We take a factor of 2 uncertainty in this study, which is conservative 
if we hold the proton composition scenario.

The differential UHE CR luminosity density, depicted in Fig.~1 above,
is of the order of $\sim10^{44}~{\rm erg}~{\rm Mpc}^{-3}~{\rm yr}^{-1}$.  
The spectral index at the sources is assumed to be
$1.5\leq s_{\rm cr}\leq2.5$. 
We remark that the differential UHE
CR luminosity density at $\sim3\times{10}^{19}$~eV that concerns us
is not too sensitive to the source spectral
index $s_{\rm cr}$, which is not easy to infer due to the degeneracy
with other parameters such as the redshift evolution 
(e.g.,~\cite{Heinze:2015hhp,AlvesBatista:2018zui}) and extragalactic
magnetic fields 
(e.g.,~\cite{Wittkowski:2017okb,Fang:2017zjf}).  This is also
consistent with the fact that the cosmogenic neutrino flux at $\sim1$~EeV does not depend on $s_{\rm cr}$ and 
the maximum CR energy at least in the proton composition case (see Figs.~6 and 7 of Ref.~\cite{Takami:2007pp}).  
Given these, one may suspect that the galactic CRs at low energies, when the leakage from galaxies
is fully taken into account, may extrapolate to the UHE range around
$3\times{10}^{19}$~eV reasonably well: in fact, the two regions can
match if the spectral index would effectively be $s_{\rm cr}\approx2.1-2.2$.

Lower-energy CR, however, if extrapolated to high energies with
the power-law index with $s_{\rm cr}\approx2.33-2.39$ as derived from AMS-02,
undershoots the UHECR luminosity density by an order of magnitude~\footnote{ 
This conclusion is unaltered if we use $s_{\rm cr}=2.25-2.33$ above 3~TeV, 
from the CREAM experiment~\cite{Yoon:2011aa}. 
The CREAM data imply that spectral indices for nuclei heavier than helium
can be harder, $s_{\rm cr}\sim2.2$ with large errors. Their fluxes 
are smaller than the proton and helium fluxes, and the UHE CRs around the ankle 
energy is dominated by light nuclei such as protons 
(even though the CRs above the ankle could consist of heavier nuclei).}. 
Matching of the low-energy and the high-energy components of CRs, requires 
tilting of the power law, say by anisotropic diffusion, reacceleration outside the galactic disc, 
or else, additional sources with a harder spectrum are needed. 
CRs cannot achieve UHE energies in remnants of ordinary ccSNe, 
so we focus on the latter two possibilities.  

For active galactic nuclei (AGN), for instance, to be a valid
candidate for such an additional agent, the requirement is 
that the CR luminosity density above TeV-PeV energies 
from additional sources overwhelm that from normal star-forming galaxies.
The luminosity density of CRs generated in AGN is unknown for sure, but we
see circumstantial evidence that the total CR luminosity density of
AGN can be on the same order of magnitude as that from normal
galaxies, as we argue in Appendix~B.  Not only AGN, we have a number of
rival candidates that would raise the luminosity density required for
UHE CRs, as also discussed in Appendices~A  and C.

The luminosity density  and the energy density
are summarised in Table~I.  Both quantities are
calculated by integrating over energies from $E=10^{18.5}$~eV to
$E=10^{20.5}$~eV.  We need to fix the slope of the spectrum: for
example, $s_{\rm cr}=2.0$ gives
$Q_{\rm uhecr}\simeq3.5\times{10}^{44}~{\rm erg}~{\rm Mpc}^{-3}~{\rm yr}^{-1}$.  
If we extrapolate this UHE CR spectrum to $E=1$~GeV, we have
$Q_{\rm uhecr}\simeq{10}^{45.3}~{\rm erg}~{\rm Mpc}^{-3}~{\rm yr}^{-1}$.  
With Eq.~(\ref{eq:uhecr}) we obtain $\Omega_{\rm uhecr}\simeq{10}^{-11.5}$, which is about
$\sim 10^{-3}$ of the total CR energy density.

\subsection{High-Energy Neutrinos}
An analysis based on a global fit of the neutrino data measured in
the IceCube experiment yields the flux~\cite{Aartsen:2015ita}
\begin{eqnarray}
E^2 \Phi_\nu&=&(6.7\pm1.2)\times{10}^{-8}~{\rm GeV}~{\rm cm}^{-2}~{\rm s}^{-1}~{\rm sr}^{-1}~\nonumber\\
&\times&{\left(\frac{E}{0.1~{\rm PeV}}\right)}^{-0.50\pm0.09},
\end{eqnarray}
where neutrinos (and antineutrinos) are added over three flavours.
This is consistent with neutrino-induced showers,
which gives the neutrino spectrum $\Phi_{\nu}\propto E^{-s}$ with
$s=2.48\pm0.08$~\cite{Aartsen:2017mau}.

The high-energy neutrino flux is also measured with upgoing muons in the
119~TeV-4.8~PeV region. Eight years of the IceCube data give~\cite{Aartsen:2017mau}
\begin{eqnarray}
E^2 \Phi_{\nu_\mu}&=&1.01^{+0.26}_{-0.23}\times{10}^{-8}~{\rm GeV}~{\rm cm}^{-2}~{\rm s}^{-1}~{\rm sr}^{-1}\nonumber\\
&\times&~{\left(\frac{E}{0.1~{\rm PeV}}\right)}^{-0.19\pm0.10}
\label{eq:upgoing}
\end{eqnarray}
per neutrino flavour.  A global fit of neutrino data gives a softer
spectrum, and this is also true in the analyses of
neutrino-induced showers and high-energy starting
events~\cite{Aartsen:2017mau}.

The arrival direction of high-energy neutrinos is
consistent with isotropic, which constrains the Galactic
contribution, and points to predominantly extragalactic nature 
of sources for IceCube neutrinos~\cite{Aartsen:2017mau,Ahlers:2013xia}.

The differential neutrino luminosity density is
\begin{eqnarray}
E{d Q_{\nu}\over dE}\approx\frac{4\pi E^2\Phi_\nu}{c\xi_z t_H}
&\simeq&{10}^{43.3\pm0.1}~{\rm erg}~{\rm Mpc}^{-3}~{\rm yr}^{-1}~\left(\frac{2}{\xi_z}\right)\nonumber\\
&\times&\left( \frac{E^2 \Phi_\nu}{3\times10^{-8}~{\rm GeV}~{\rm cm}^{-2}~{\rm s}^{-1}~{\rm sr}^{-1}}\right), \,\,\,\,\,\,\,\,\cr
&&
\label{nubudget}
\end{eqnarray}
where $\xi_z=t_H^{-1}\int\frac{dz}{1+z}|\frac{dt}{dz}|(Q_{\nu}(z)/Q_{\nu}) \approx 2$ 
is the correction factor for the integration over redshift with the star-formation history we adopt~\cite{Madau:2014bja}. 
We note that the line-of-sight integral in the diffuse flux calculation is 
dominated by sources at $z\sim1$. 
The neutrino luminosity densities derived from the IceCube data are also presented in Fig.~1 above. 
Eq.~(\ref{eq:upgoing}) is multiplied by a factor of 3 in the figure to take into account the other flavors.

The global analysis of the 10-100 TeV neutrino data 
gives  $Q_\nu\simeq1.8\times{10}^{44}~{\rm erg}~{\rm Mpc}^{-3}~{\rm yr}^{-1}~(2/\xi_z)$, and 
the upgoing muon neutrino analysis focusing on $\gtrsim 100$ TeV leads to
$Q_\nu\simeq5.6\times{10}^{43}~{\rm erg}~{\rm Mpc}^{-3}~{\rm yr}^{-1}~(2/\xi_z)$ (see Table~I).  
The latter is by a factor of 3 larger because of the higher neutrino flux in the 10-30
TeV range.  If the neutrino spectrum is extended to lower energies
with $s\sim2.5-3.0$, both $Q_{\nu}$ and $\Omega_{\nu}$ will be even larger.

High-energy neutrinos are produced by CRs through inelastic $pp$ or
$p\gamma$ interactions, yielding the energy flux
(e.g.,~\cite{Murase:2015xka,Waxman:1998yy}),
\begin{equation}
E_\nu^2 \Phi_\nu^{(pp/p\gamma)}\approx\frac{c \xi_z t_H}{4 \pi}\frac{3K}{4(1+K)}{\rm min}[1,f_{\rm meson}]E_p\frac{d Q_{\rm excr}}{d E_p},
\label{neuflux}
\end{equation}
where $K=1$ or 2 for $p\gamma$ or $pp$ interactions, respectively, and
$f_{\rm meson}$ is the the effective optical depth for the meson production. 
With $E_\nu^2 \Phi_\nu =E_\nu^2 \Phi_\nu^{(pp/p\gamma)}$ we are led to a constraint on the CR luminosity 
density at energies relevant to IceCube neutrinos:
\begin{eqnarray}
E_p{d Q_{\rm excr}\over dE_p}&\approx&4.4\times {10}^{43}~{\rm erg}~{\rm Mpc}^{-3}~{\rm yr}^{-1}~\frac{[4(1+K)/6K]}{{\rm min}[1,f_{\rm meson}]}\nonumber\\
&\times&~\left(\frac{2}{\xi_z}\right)\left(\frac{E_\nu^2\Phi_\nu}{3\times10^{-8}~{\rm GeV}~{\rm cm}^{-2}~{\rm s}^{-1}~{\rm sr}^{-1}}\right). \,\,\,\,\,\,\,\,\cr
&&
\label{nulimit}
\end{eqnarray}
Imposing ${\rm min}[1,f_{\rm meson}]\leq1$ gives a lower limit on the
differential CR luminosity density~\cite{Katz:2013ooa}.  At a
few PeV energies, this is on the correct order of magnitude of the CR
luminosity densities extrapolated from either the GeV-TeV range or UHE range.

\subsection{High-Energy Gamma Rays}
The {\it Fermi} Large Area Telescope has identified a large number of extragalactic gamma-ray sources 
that consist of blazars, radio galaxies, and actively star-forming galaxies, but also it gives the 
``extragalactic gamma-ray background (EGB)'' above
0.1~GeV~\cite{Ackermann:2014usa},  as
\begin{eqnarray}
E_\gamma^2\Phi_\gamma&=&(1.48\pm0.09)\times{10}^{-6}~{\rm GeV}~{\rm cm}^{-2}~{\rm s}^{-1}~{\rm sr}^{-1}~\nonumber\\
&\times&{\left(\frac{E}{0.1~{\rm GeV}}\right)}^{-0.31\pm0.02}.
\label{EGB}
\end{eqnarray}
At $\gtrsim50$~GeV energies one considers that the EGB is
dominated by unresolved point sources, mainly of
blazars~\cite{TheFermi-LAT:2015ykq,Zechlin:2015wdz,Lisanti:2016jub}.
The remaining background, including both unresolved point sources 
and diffuse component, is referred as the ``isotropic diffuse gamma-ray background
(IGRB)''. The IRGB flux up to $\sim1$~TeV is~\cite{Ackermann:2014usa}
\begin{eqnarray}
E_\gamma^2\Phi_\gamma&=&(0.95\pm0.08)\times{10}^{-6}~{\rm GeV}~{\rm cm}^{-2}~{\rm s}^{-1}~{\rm sr}^{-1}~\nonumber\\
&\times&{\left(\frac{E}{0.1~{\rm GeV}}\right)}^{-0.32\pm0.02}.
\label{IGRB}
\end{eqnarray}

As in Eq.~(\ref{nubudget}) the gamma-ray background flux can be translated into the differential gamma-ray luminosity density 
\begin{eqnarray}
E{d Q_{\gamma}\over dE}\approx\frac{4\pi E^2\Phi_\gamma}{c\xi_z t_H}
&\simeq&{10}^{43.86\pm0.05}~{\rm erg}~{\rm Mpc}^{-3}~{\rm yr}^{-1}~\left(\frac{2}{\xi_z}\right)\nonumber\\
&\times&\left( \frac{E^2 \Phi_\gamma}{10^{-7}~{\rm GeV}~{\rm cm}^{-2}~{\rm s}^{-1}~{\rm sr}^{-1}}\right), \,\,\,\,\,\,\,\,
\label{gammabudget}
\end{eqnarray}
which gives $Q_\gamma\simeq3.3\times{10}^{45}~{\rm erg}~{\rm Mpc}^{-3}~{\rm yr}^{-1}~(2/\xi_z)$ for the total EGB 
and $Q_\gamma\simeq2.1\times{10}^{45}~{\rm erg}~{\rm Mpc}^{-3}~{\rm yr}^{-1}~(2/\xi_z)$ for the IGRB.
The gamma-ray luminosity densities derived from the EGB and IGRB are also shown in Fig.~1.

Gamma rays are produced by leptonic processes such as the inverse-Compton radiation in addition to hadronic processes 
of $pp$ and $p\gamma$ interactions. For the hadronic component 
the generated energy flux of gamma rays from neutral pion decay is, similarly to Eq.~(\ref{neuflux}), 
\begin{equation}
E_\gamma^2 \Phi_\gamma^{(pp/p\gamma)}\approx\frac{c \xi_z t_H}{4 \pi} \frac{1}{1+K}{\rm min}[1,f_{\rm meson}]E_p\frac{d Q_{\rm excr}}{d E_p}.
\end{equation}
The fact that the EGB and IGRB receive both leptonic and hadronic contributions means
$E_\gamma^2 \Phi_\gamma^{(pp/p\gamma)}\leq E_\gamma^2 \Phi_\gamma$,
which leads to an upper limit 
\begin{eqnarray}
E_p{d Q_{\rm excr}\over dE_p}&\lesssim&2.2\times {10}^{44}~{\rm erg}~{\rm Mpc}^{-3}~{\rm yr}^{-1}~\frac{[(1+K)/3]}{{\rm min}[1,f_{\rm meson}]}\nonumber\\
&\times&~\left(\frac{2}{\xi_z}\right)\left(\frac{E_\gamma^2\Phi_\gamma}{10^{-7}~{\rm GeV}~{\rm cm}^{-2}~{\rm s}^{-1}~{\rm sr}^{-1}}\right). \,\,\,\,\,\,\,\,\cr
&&
\label{gammalimit}
\end{eqnarray}

Eq.~(\ref{gammalimit}) is valid for gamma rays in the GeV range.  Very
high-energy gamma rays with $E_\gamma\gtrsim0.1$~TeV, however,
interact with the extragalactic background light (EBL) and cosmic
microwave background (CMB), causing electron-positron pair creation.
The generated electrons and positrons lose their energies through
inverse-Compton scattering, producing electromagnetic
cascades, whose resulting gamma rays appear in the MeV-TeV region.  The
spectrum of gamma rays may be characterised effectively with
$G_\gamma$, the ratio of the differential energy spectrum of cascade
gamma rays at specific energy $E_\gamma$ to the integrated energy
spectrum of injected gamma rays, electrons and positrons
~\cite{Berezinsky:1975zz,Murase:2012df}, such that
$E_\gamma(dG_\gamma/dE_\gamma)\equiv E_\gamma^2\phi_\gamma^{(\rm
  cas)}/ \int dE_\gamma \phi_\gamma^{(\rm em)}$,
\begin{equation}
\frac{dG_\gamma}{dE_\gamma}\propto 
\left\{ \begin{array}{ll}
{(E_{\gamma}/E_{\gamma}^{\rm br})}^{-1/2}
& \mbox{($E_\gamma \leq E_{\gamma}^{\rm br}$)}\\
{(E_{\gamma}/E_{\gamma}^{\rm br})}^{1-\beta} 
& \mbox{($E_{\gamma}^{\rm br} < E_\gamma \leq E_{\gamma}^{\rm cut}$)}.
\end{array} \right. 
\end{equation} 
where $dG_{\gamma}/dE_\gamma$ is normalised as $\int d E_\gamma \,
(dG_{\gamma}/dE_\gamma)=1$;  the break energy $E_{\gamma}^{\rm br}\approx(4/3)
{({E'}_\gamma^{\rm cut}/m_e c^2)}^2 \varepsilon_{\rm CMB} \simeq
0.034~{\rm GeV}~{(E_{\gamma}^{\rm cut}/0.1~{\rm TeV})}^2 {[(1+z)/2]}^{2}$ 
with $\varepsilon_{\rm CMB}$ the CMB energy, 
$E_{\gamma}^{\rm cut}$ is the cutoff due to EBL and the spectral index $\beta \sim 2$. 
Averaging over redshifts, 
$E_\gamma^{\rm br}(d\bar{G}_{\gamma}/dE_\gamma)|_{E_\gamma^{\rm br}}\sim{(2+\ln(E_{\gamma}^{\rm cut}/E_{\gamma}^{\rm br}))}^{-1}\sim0.1$,
which is consistent with a numerical calculation~\cite{Murase:2012df}.
The detail of initial spectra is unimportant for they are largely washed out.

Electromagnetic cascades convert the bolometric electromagnetic energy, when input at 
sufficiently high energies, to lower-energy gamma rays in the {\it Fermi} range. 
The resulting gamma-ray background flux is~\cite{Murase:2012df}
\begin{eqnarray}
E_\gamma^2 \Phi_{\gamma}^{(\rm cas)}\approx\frac{c\xi_z t_H}{4 \pi} Q_\gamma^{(\rm em)}E_\gamma\frac{d\bar{G}_{\gamma}}{dE_{\gamma}}, 
\label{eq:cas}
\end{eqnarray}
where $Q_{\gamma}^{(\rm em)}\approx 4\pi\int dE_\gamma \Phi_\gamma^{(\rm em)}/(c\xi_zt_H)$ is the electromagnetic 
luminosity density due to the photomeson production and the Bethe-Heitler pair production. 
Using Eqs.~(\ref{IGRB}) and (\ref{eq:cas}), one obtains~\cite{Murase:2012df,Decerprit:2011qe}
\begin{eqnarray}
Q_{\gamma}^{(\rm em)}&\lesssim&8\times{10}^{44}~{\rm erg}~{\rm Mpc}^{-3}~{\rm yr}^{-1}~{(2/\xi_z)}\nonumber\\
&\times&\left(\frac{E_\gamma^2\Phi_\gamma}{10^{-7}~{\rm GeV}~{\rm cm}^{-2}~{\rm s}^{-1}~{\rm sr}^{-1}}\right), 
\label{eq:gamlim2}
\end{eqnarray}
which is taken as an upper limit on the UHE CR luminosity density, 
since the effective optical depth is $\gtrsim1$ for the photomeson
and the Bethe-Heitler pair production of UHE CRs above $E\gtrsim{10}^{18.5}$~eV.

\section{Energy density of cosmic radiations}
With the generation rate density $Q_{\rm cr}$, the CR energy density is 
\begin{equation}
E{dU_{\rm cr}\over dE}=E{d\Omega_{\rm cr}\over dE} (\varrho_{\rm crit}c^2)\approx\int^{t_{\rm surv}} \frac{dt}{1+z} E' {dQ_{\rm cr}\over dE},
\end{equation}
where $E'=(1+z)E$ and $t_{\rm surv}\approx{\rm min}[t_H,t_{\rm loss}]$ is the survival time.
In the absence of energy losses during the Hubble time, 
\begin{eqnarray}
E{d\Omega_{\rm cr}\over dE}&\approx&\frac{\xi_z  t_H}{\varrho_{\rm crit}c^2}E{dQ_{\rm cr}\over dE}\nonumber\\
&\simeq&{10}^{-9.1\pm0.3}~{\left(\frac{E}{10~{\rm GeV}}\right)}^{2-s_{\rm cr}}(\xi_z/2),
\end{eqnarray}
where $\xi_z=t_H^{-1}\int\frac{dz}{1+z}|\frac{dt}{dz}|(Q_{\rm cr}(z)/Q_{\rm cr}) \approx 2$ is introduced as in Sec.~II. 
Integrating over energies, we obtain the global CR energy density in units of the critical energy density, $\varrho_{\rm crit}c^2$,
\begin{equation}
\Omega_{\rm cr}=U_{\rm cr}/(\varrho_{\rm crit}c^2)\simeq{10}^{-8.3\pm0.3}~(\xi_z/2), 
\end{equation}
in agreement with Ref.~\cite{Fukugita:2004ee}.

Fig.~2 shows the cosmic energy density of CRs per logarithmic energy,
$d\Omega_{\rm cr}/d\ln E$, translated from Fig.~1.  
The upper (grey) shaded region is the global CR energy density, corresponding to the upper curves in Fig.~1.
We also show the energy density of CRs that reside in the galactic disc, discarding the escaped CRs, 
with (brown) shades in the lower part of the figure. 
If, as in our archetype argument, the CRs generated in {\rm a galaxy} escape, 
then the energy density of confined CRs amounts to $t_{\rm esc}(E)/(\xi_zt_H)$ 
times the above estimate. 
Here it is assumed that Eq.~(\ref{eq:escapetime}), given for the Milky Way, 
represents the typical escape time of galactic CRs.

In reality, low-energy CRs may undergo adiabatic energy
losses in the expanding wind~\cite{Lacki:2013ata} and  
inelastic $pp$ collisions, in particular, in actively star-forming galaxies 
(starburst galaxies), in which the $pp$ cooling time is likely to be shorter than the
advection and the escape times~\cite{Loeb:2006tw,Thompson:2006np}. 
Thus, whereas the above estimate, corresponding to the upper curve in Fig.~2,
may be valid in normal galaxies, this $\Omega_{\rm cr}$ we obtained is taken, for sure, 
as an upper limit on the current energy density of CRs with stellar origins. 

For extragalactic radiations, including UHE CRs, neutrinos, and gamma rays,
the differential global energy density is written in terms of the observed energy density as,
\begin{equation}
E{d\Omega\over dE}=E{dU\over dE}{1\over\varrho_{\rm crit}c^2}={4\pi E^2\Phi\over c}{1\over\varrho_{\rm crit}c^2}. 
\end{equation}
The UHE CR energy density on Earth is measured accurately, but the global energy density
depends on the transition from Galactic to extragalactic CRs.  In Fig.~2, 
the upper curve of the UHE CR shade region is taken from
Ref.~\cite{Fang:2017zjf} which uses the mixed composition model with
the proton dominance around the ankle energy.  
For $E\gtrsim10^{18.5}$~eV this model has the spectrum that is close to
the flux given by Eq.~(\ref{eq:uhecr}).  
The lower curve of the (green) shade region is based on the classical proton ankle model with
$s_{\rm sc}=2.0$~\cite{Decerprit:2011qe,Murase:2016gly}, which can be
taken as a more conservative estimate on the UHE CR energy density.  Note
that both curves agree with each other above $E\sim10^{19}$~eV,
where the extragalactic component is dominant.

High-energy neutrino observations serve as a probe for the 
differential luminosity and energy density of CRs. For a given spectral index
we obtain the total luminosity density: for $s=2.5$ we have
$Q_{\nu}\simeq{10}^{44.3\pm0.1}~{\rm erg}~{\rm Mpc}^{-3}~{\rm
  yr}^{-1}~(2/\xi_z)$ and $\Omega_{\nu}\simeq{10}^{-10.7\pm0.1}$ above
25~TeV (see Table~I).  This yields
$Q_{\rm excr}\gtrsim{10}^{44.6\pm0.1}~{\rm erg}~{\rm Mpc}^{-3}~{\rm
  yr}^{-1}~(2/\xi_z)$ and $\Omega_{\rm excr}\gtrsim{10}^{-10.4\pm0.1}$
above $E_p\sim0.5$~PeV, assuming inelastic $pp$ interactions being the
dominant process.  


As indicated in Fig.~3 below, the GeV-TeV CR spectrum extrapolated to
PeV energies can be compatible with the soft CR spectrum indicated by
the 10-100~TeV neutrino data if the meson production is fully
efficient for PeV CRs, i.e. the effective optical depth $f_{\rm meson}\gtrsim1$. 
On the other hand, $\Omega_{\rm cr}\gg \Omega_\gamma$ means that the CR energy
density in the GeV-TeV range needed to explain GeV-TeV gamma rays 
overshoots the observed CR roughly by a factor of $3-10$ (see Fig.~3).  
This implies that the gamma-ray production for GeV-TeV CRs should not be fully
efficient (i.e.  $f_{\rm meson}\ll1$) and/or the hadronic gamma rays
are attenuated by interactions with ambient matter or radiation.  A
comparison of the IceCube and {\it Fermi} data implies that the CR
accelerators are hidden when viewed in GeV-TeV gamma rays~\cite{Murase:2015xka}.

\begin{figure}[t]
\includegraphics[width=3.50in]{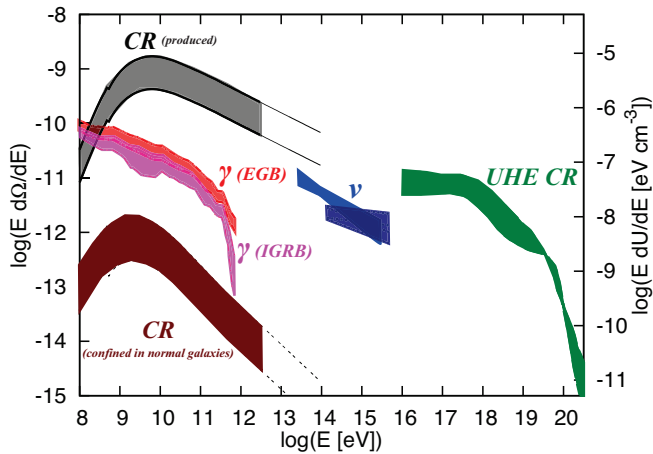}
\caption{
Local ($z=0$) energy densities of cosmic particles (in unit of $\varrho_{\rm crit}c^2$ for the left axis). 
We ignore possible energy losses of CRs injected in the past. 
Unavoidable energy losses of UHE CRs are taken into account, and we show both proton ankle and mixed composition models. 
The energy density of CRs confined in normal galaxies is indicated assuming $t_{\rm esc}=15-100$~Myr at $R=1$~GV.
The softening of the EGB and IGRB spectra is caused by the attenuation by the extragalactic background light. 
\label{fig2}
}
\vspace{-1.\baselineskip}
\end{figure}

%
\section{Discussion}
\subsection{Cosmic-Ray Reacceleration in Star-Forming Galaxies}
The CR energy generation rate density in the GeV-TeV range, when leaked
CRs are included, is not much different from that in the UHE range,
across energies 10 orders of magnitude apart. 
This may point towards the idea that CRs may be reaccelerated by the Fermi acceleration in the vicinity of galaxies, say, in
superbubbles or galactic winds driven by the star-formation, whose power-law index can be
$s_{\rm cr}\sim2$ or smaller~\cite{1987ApJ...312..170J,Volk:2004uc}.

The velocity of galactic winds driven by the star-formation can be 
$V_s\sim300-1000~{\rm km}~{\rm s}^{-1}$~\cite{1990ApJS...74..833H,2011ApJ...733..101G} 
(see also reviews, e.g.,~\cite{Veilleux:2005ia,2017arXiv170109062H}), and the 
termination shock radius would be $R_s\sim5-50$~kpc (e.g.,~\cite{Lacki:2013ata}). 
The Fermi bubbles, which are seen in gamma rays, could be attributed to the termination 
shock formed by starburst winds. 
With a magnetic field of $B\sim5~\mu {\rm G}$~\cite{Su:2010qj,Lacki:2013ata}, the maximum energy of CRs is 
\begin{eqnarray}
E_{\rm max}&\approx& \frac{3}{20}ZeB(V_s/c)R_s\nonumber\\
&\sim&{10}^{16}~{\rm eV}Z~\left(\frac{B}{5~\mu\rm G}\right)\left(\frac{V_s}{1000~\rm km~s^{-1}}\right)\left(\frac{R_s}{5~\rm kpc}\right),\cr
&&
\end{eqnarray}
which means that galactic CRs in the halo can be reaccelerated 
up to or even beyond the knee energy $\approx3-4$~PeV.  
The Galactic CR spectrum may be explained by the reacceleration 
of cosmic rays originating from supernovae. 

The actively star-forming galaxies, often referred to as starburst galaxies, 
may have a higher wind velocity of
$V_s\sim1000-2000~{\rm km}~{\rm s}^{-1}$~\cite{1990ApJS...74..833H,2011ApJ...733L..16S} 
and a larger magnetic field of $B\sim0.1-1$~mG~\cite{Thompson:2006is}, with which the maximum
energy could reach $E_{\rm max}\sim{10}^{20.5}(Z/26)~$~eV.  
This value is significantly higher than that for the Milky Way.
Thus, if galactic CRs can be reaccelerated in the galactic halo with a hard spectral index of
$s_{\rm cr}\sim2.0-2.2$, it is possible that the global CR generation rate density in the GeV-TeV
range could match that in the UHE range, up to one order of
magnitude across the energy difference extending 10 orders of magnitude.
In this scenario, the highest-energy CRs are largely nuclei and the correlation between UHE CRs 
and starburst galaxies can be expected.

\subsection{Active Galactic Nuclei and Structure Formation Shocks}
As detailed in Appendix~B, recent observations revealed that luminosity
densities of nonthermal gamma-ray emissions from AGN, including blazars, lie in the range
$Q_{\gamma}^{\rm AGN}\sim3\times({10}^{44}-{10}^{46})~{\rm erg}~{\rm
  Mpc}^{-3}~{\rm yr}^{-1}$.  The observed X-ray and gamma-ray
emissions are attributed to synchrotron and inverse-Compton radiation
from relativistic electrons.  
We expect that ions may also be accelerated by the Fermi acceleration mechanism, 
so that the CR luminosity density for AGN may be comparable to or even larger than
that for star-forming galaxies,
$Q_{\rm cr}\simeq2\times{10}^{46}~{\rm erg}~{\rm Mpc}^{-3}~{\rm yr}^{-1}$, 
as well as $Q_{\gamma}^{\rm AGN}$.  The gamma-ray
luminosity densities for blazars and radio galaxies are similar to
that for starburst galaxies, $Q_{\gamma}^{\rm SBG}\sim3\times{10}^{44}
~{\rm erg}~{\rm Mpc}^{-3}~{\rm yr}^{-1}$, which is consistent with the idea that EGB
consists of multiple components.

CRs may be accelerated by shocks in hierarchical structure formation, 
which induces accretion shocks, cluster and galaxy mergers.  
The integrated CR luminosity densities associated with these shocks are 
$Q_{\rm cr}\sim3\times({10}^{45}-{10}^{46})
~{\rm erg}~{\rm Mpc}^{-3}~{\rm yr}^{-1}$ (see Appendix~C).  
This is also comparable to that for star-forming galaxies.

\subsection{Consistency among High-Energy Cosmic Radiations?}

In summary, Fig.~3 reiterates the energy generation rate density of
CRs.  In this figure energies of high-energy neutrinos and gamma rays
are those of primary CRs.  IceCube neutrinos
give constraints on the CR luminosity density in the 1$-$100~PeV
range.  In particular, for $s_{\rm cr}\sim2.0-2.3$, CR spectra
extrapolated from UHE energies can match the CR luminosity needed to explain
the IceCube data, if the efficiency to produce neutrinos is
sufficiently high (i.e. min$[1,f_{\rm meson}]\sim0.03-1$); see Fig.~3
and Eq.~(\ref{nulimit}). The IGRB gives an upper limit on the CR
luminosity density for a given $f_{\rm meson}$ (see
Eq.~\ref{gammalimit}; Fig.~3 displays the case for
$f_{\rm meson}\geq1$). This upper limit becomes weaker
if $f_{\rm meson}<1$. 
The figure indicates that hard CR spectra with $s_{\rm cr}\sim2.0-2.2$ 
can explain the UHE CR and neutrino fluxes without violating the 
upper limits from the IGRB; see also Ref.~\cite{Murase:2013rfa}.

That UHE CR flux around the ankle energy is compatible with those of
0.1~PeV neutrinos and 0.1~TeV gamma rays (see Fig.~2 above). This
implies that the UHE CR luminosity density is comparable to the CR
energy generation rate density to explain the high-energy neutrino
background in the sub-PeV range and the IGRB in the sub-TeV range (see
Fig.~3).  This fact naturally leads to an idea that UHE CRs,
neutrinos, and gamma rays may share a common origin.
As an example, let us consider the CR luminosity density given by
Eq.~(\ref{uhebudget}), which yields
$E(dQ_{\rm uhecr}/dE)\approx0.75\times{10}^{44}~{\rm erg}
~{\rm Mpc}^{-3}~{\rm yr}^{-1}$ for $s_{\rm sr}\sim2$.  From
Eq.~(\ref{nulimit}) the IceCube neutrino flux can be explained if the
following condition is met:
\begin{equation}
{\rm min}[1,f_{\rm meson}]\sim0.6~[4(1+K)/6K]{(2/\xi_z)}.
\end{equation}
With $K=2$ and $\xi_z=2$ this means $f_{\rm mes}\sim0.6$ (cf. Fig.~1 of Ref.~\cite{Murase:2016gly}). 
Such high-energy neutrino sources also satisfy Eq.~(\ref{gammalimit}), since 
\begin{equation}
{\rm min}[1,f_{\rm meson}]\lesssim3~[(1+K)/3]{(2/\xi_z)}. 
\end{equation}
This implies that gamma-ray flux from $pp$ interactions is
$E_{\gamma}^2\Phi_\gamma\sim2\times{10}^{-8}~{\rm GeV}
~{\rm cm}^{-2}~{\rm s}^{-1}~{\rm sr}^{-1}$ with $f_{\rm meson}\sim0.6$.
Including the cascade component that would enhance the flux by a factor of 2, we have
$E_{\gamma}^2\Phi_\gamma\sim4\times{10}^{-8}~{\rm GeV}
~{\rm cm}^{-2}~{\rm s}^{-1}~{\rm sr}^{-1}$, as consistent with a
numerical calculation~\cite{Murase:2013rfa}. UHE CRs also produce
cosmogenic gamma rays during their propagation in intergalactic space.
At $E\gtrsim10^{18.5}$~eV, $\sim65-100$~\% of the UHE CR proton 
is converted to the electromagnetic energy.
Thus, one finds that the IGRB constraint on $Q_{\gamma}^{(\rm em)}$ 
(see Eq.~\ref{eq:gamlim2}) is readily met, and the corresponding
cosmogenic gamma-ray flux is
$E_{\gamma}^2\Phi_\gamma\sim4\times{10}^{-8}~{\rm GeV}
~{\rm cm}^{-2}~{\rm s}^{-1}~{\rm sr}^{-1}$.  Summing up contributions
originating from $pp$ interactions in the source and $p\gamma$
interactions in intergalactic space, the total gamma-ray flux is
$E_{\gamma}^2\Phi_\gamma\sim8\times{10}^{-8}~{\rm GeV}~{\rm cm}^{-2}
~{\rm s}^{-1}~{\rm sr}^{-1}$ (at $\sim10$~GeV), which matches the non-blazar
component of the EGB~\cite{Ackermann:2014usa,Lisanti:2016jub}.  
(We note that the sub-TeV spectrum of the EGB is affected by gamma-ray 
attenuation due to the extragalactic background light.)
This result also agrees with Fig.~1 of Ref.~\cite{Murase:2016gly}, and
the CR spectral index needs to be $s_{\rm cr}\lesssim2.1$ in simple 
power-law models (see also Ref.~\cite{Fang:2017zjf}, in which such a hard CR 
index is effectively achieved for CRs leaving their accelerators).  
A hard CR spectrum extrapolated from UHE energies downwards can 
simultaneously account for high-energy neutrinos and gamma rays.
UHE CRs may come from either reacceleration of galactic CRs in the halo 
or additional source population. 

\begin{figure}[t]
\includegraphics[width=3.50in]{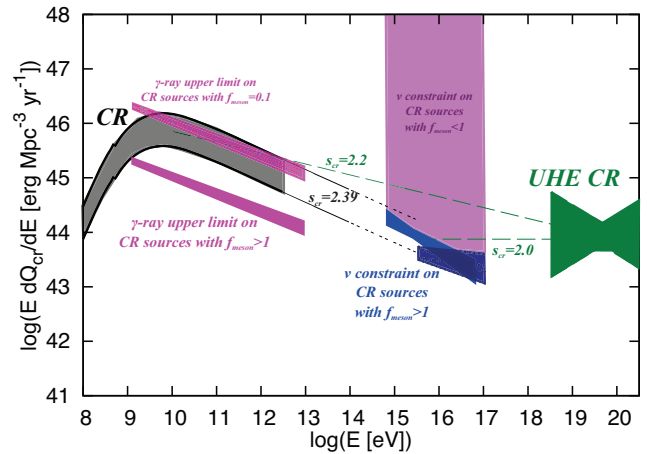}
\caption{Similar to Fig.~1, but neutrino and gamma-ray luminosity densities 
are converted into CR energy generation rate densities assuming that they are 
produced by inelastic $pp$ interactions. 
The high-energy neutrino background gives constraints on the CR energy generation rate density, 
while the IGRB is taken as an upper limit for a given value of the effective optical depth to meson
production ($f_{\rm meson}$): the cases for $f_{\rm meson}\geq1$ and $f_{\rm meson}=0.1$ are displayed.  
\label{fig3}
}
\vspace{-1.\baselineskip}
\end{figure}

%
\section{Summary}
We have studied energy generation rate densities and energy densities
of nonthermal cosmic particles, using the recent experiments of CRs at
both GeV-TeV region and ultrahigh energies, and in addition high-energy 
neutrinos and gamma rays.  
The total energy density of GeV-TeV CRs is $\Omega_{\rm cr}\simeq10^{-8.3}$. 
Extrapolation to UHE energies, across the energy range of 10 orders of magnitude, 
gives, if we would adopt the local CR spectral index derived from the AMS-02 experiment 
at GeV-TeV energies, nominally a value smaller than the observed UHE CR 
luminosity density by a factor of $\sim30$.  
That is, the UHE CRs do not quite match the extrapolation from the GeV-TeV CRs 
if the global CR spectral index is the same as the local one, $s_{\rm cr}\approx2.3-2.4$.
The matching requires somewhat harder spectra, say $s_{\rm cr}\approx2.1-2.2$.

The gap between GeV-TeV CRs and UHE CRs leaves a room for
reacceleration of GeV-TeV CRs, or additional components contributing
to the high-energy CRs. In fact, the energy stored in circumgalactic
and cluster warm matter is $\Omega\simeq10^{-8.0\pm0.2}$~\cite{Fukugita:2004ee}.  
This is larger than the total energy density of CRs, so that it
can be used to reaccelerate CRs.  Alternatively, we can think of a
number of candidate sources that would fill the gap, such as AGN
and/or similar active objects, or some violent transients that have
comparable energies when integrated over time. We argue that it is
energetically possible that any of such sources could fill the gap.
For the moment we are unable to distinguish between the two possibilities,
or single out some candidate sources.

It is interesting to observe that high-energy cosmic neutrinos and
gamma rays indicate the relevant luminosity densities that are,
roughly speaking, comparable to that of the UHE CRs.  In fact, the
observed high-energy neutrino flux is consistent with extragalactic
CRs, but the CRs needed to produce gamma rays, at an estimate
of the maximum meson production efficiency, should be smaller than GeV-TeV CRs.  
The gamma-ray production may be suppressed effectively, say, by small
effective optical depths for meson production and/or by large optical
depths to two photon annihilation.  As a global picture our energetics
argument is suggestive of a common origin of UHE CRs, high-energy
neutrinos and gamma rays~\cite{Fang:2017zjf}.


\medskip
\begin{acknowledgments}
  We thank Kfir Blum and Yutaka Ohira for discussions
  on cosmic-ray propagation and interpretations of the spectral hardening.
  We also thank Eli Waxman for discussion as to the origin of IceCube
  neutrinos below 100~TeV.  The work of KM is
  supported by the Alfred P. Sloan Foundation and the U.S. National
  Science Foundation (NSF) under grants NSF Grant No. PHY-1620777.  He
  also thanks Institute for Advanced Study for its hospitality during
  the work.  MF thanks Hans B\"ohringer and late Yasuo Tanaka for the
  hospitality at the Max-Planck-Institut f\"ur Extraterrestrische
  Physik and also Eiichiro Komatsu at Max-Planck-Institut f\"ur
  Astrophysik, in Garching. He also wishes his thanks to Alexander von
  Humboldt Stiftung for the support during his stay in Garching, and
  Monell Foundation in Princeton.  He received in Tokyo a Grant-in-Aid
  (No. 154300000110) from the Ministry of Education. Kavli IPMU is
  supported by World Premier International Research Centre Initiative
  of the Ministry of Education, Japan.
\end{acknowledgments}

\vskip15mm
\centerline{\bf Appendix~~Luminosity Density of Various Sources}
\vskip7mm

We discuss here a number of candidate sources that may contribute
significantly to CRs, with the fact in mind that robust estimates for
their CR energy generation rate density are difficult with the present knowledge and
hence they remain as order of magnitude estimates. 
  
We write the luminosity density $Q(z)$ at $z$ of some astrophysical source
\begin{equation}
Q(z)=\int dL\,L\frac{dn_s}{dL}(L,z),
\end{equation}
where $n_s=\int dL\, (dn_s/dL)$ is the number density of the source
per comoving volume.  For the transient source we write the
luminosity density,
\begin{equation}
Q(z) = \int d {\mathcal E} \, {\mathcal E} \frac{d\rho_s}{d{\mathcal E}}({\mathcal E},z),
\end{equation}
where ${\mathcal E}$ is the energy and $\rho_s=\int d{\mathcal E}\, (d\rho_s/d{\mathcal E})$ is 
the number of the transient sources per comoving volume per time (rate density).  
We denote their local luminosity density as $Q$.

\begin{table*}[t]
\begin{center}
\caption{The luminosity density $Q$ in units of ${\rm erg}~{\rm
    Mpc}^{-3}~{\rm yr}^{-1}$ and the rate density $\rho$ in units of
  ${\rm Mpc}^{-3}~{\rm yr}^{-1}$ or number density $n$ in units of ${\rm
    Mpc}^{-3}$, as discussed in the text: core-collapse supernova
  (ccSNe), hypernovae (HN), double neutron star mergers (DNS), 
  gamma-ray bursts (GRB), low-luminosity gamma-ray bursts (LL GRB), 
  tidal disruption events (TDE), BL Lac objects (BL Lac), flat-spectrum
  radio quasars (FSRQ), radio galaxies (RG), and accretion or merger
  shocks due to the cosmic structure formation (Accr/Mger).  CR
  luminosity densities are those estimated from kinetic luminosity
  densities assuming $\epsilon_{\rm cr}=0.1$. For GRB and TDE the
  $\gamma$-ray luminosity densities are based on the BATSE and the
  {\it Swift} band observations. For BL Lac and FSRQ data are taken
  from {\it Fermi} Large Area Telescope observations. 
\label{tb1}
}
\scalebox{1.0}{
\begin{tabular}{|c|c|c|c|c|c|c|}
\hline         & ccSN (CR) & HN (CR) & DNS (CR) & GRB ($\gamma$) & LL GRB ($\gamma$) & TDE  ($\gamma$)\\
\hline $Q$ [${\rm erg}~{\rm Mpc}^{-3}~{\rm yr}^{-1}$] & ${10}^{46.6}$ & ${10}^{45.5}$ & ${10}^{44.5}$ & ${10}^{43.6}$ & ${10}^{43.5}$ & ${10}^{43.5}$ \\
\hline $\rho$ [${\rm Mpc}^{-3}~{\rm yr}^{-1}$] & ${10}^{-4}$ & ${10}^{-5.5}$ & ${10}^{-5.8}$ & ${10}^{-9}$ & ${10}^{-6.5}$ & ${10}^{-10.5}$ \\
\hline
\hline         &  SBG ($\gamma$) & AGN (X) & BL Lac ($\gamma$) & FSRQ  ($\gamma$) & RG ($\gamma$) & Accr/Mger (CR)\\
\hline $Q$ [${\rm erg}~{\rm Mpc}^{-3}~{\rm yr}^{-1}$] & ${10}^{44.5}$ & ${10}^{46.3}$ & ${10}^{45.4}$ & ${10}^{44.3}$ & ${10}^{44.6}$ & ${10}^{46.5}$ \\
\hline $n$  [${\rm Mpc}^{-3}$] & ${10}^{-4}$ & ${10}^{-4}-{10}^{-3}$ & ${10}^{-7}-{10}^{-6.5}$ & ${10}^{-9}-{10}^{-8}$ & ${10}^{-5}-10^{-4}$ & ${10}^{-6}-10^{-5}$ \\
\hline
\end{tabular}
}
\end{center}
\end{table*}

\subsection{Energetic Transients in Star-Forming Galaxies}
Our result in the main text (Table~I above) shows the luminosity density of galactic CRs,
\begin{equation}
Q_{\rm cr}\sim4\times{10}^{46}~{\rm erg}~{\rm Mpc}^{-3}~{\rm yr}^{-1}.
\label{crbudget}
\end{equation}
This is compared with CRs to be generated in ccSNe showing that the
two values are consistent; the bulk of CR can be ascribed to ccSNe.

In addition to ordinary ccSNe, rare but more energetic transients may
occur, in particular, in star-forming galaxies, such as broadline Type Ibc supernovae, 
often called hypernovae with ${\mathcal E}_{\rm HN}\sim10^{52}$~erg.
The broadline Type Ibc rate is approximately $\rho_{\rm HN}\sim3000~{\rm Gpc}^{-3}~{\rm yr}^{-1}$~\cite{Guetta:2006gq},
which is $0.032\pm0.007$ times the ccSN rate.
The CR luminosity density of hypernovae is then
\begin{equation}
Q_{\rm cr}^{\rm HN}\sim3\times{10}^{45}~{\rm erg}~{\rm Mpc}^{-3}~{\rm yr}^{-1}~\epsilon_{\rm cr,-1}{\mathcal E}_{\rm HN,52}. 
\end{equation}
If $s_{\rm cr}\sim2$ and the maximum energy
$E_{\rm max}=10^{20.5}$~eV, the differential UHE CR
luminosity density $E(dQ_{\rm cr}^{\rm HN}/dE)\sim1\times{10}^{44}~{\rm erg}~{\rm Mpc}^{-3}~{\rm yr}^{-1}$ 
is comparable to $E(dQ_{\rm uhecr}/dE)\approx0.75\times{10}^{44}~{\rm erg}~{\rm Mpc}^{-3}~{\rm yr}^{-1}$ 
which is the UHE CR luminosity density given by Eq.~(\ref{uhebudget}).

One may suspect that Type Ia supernovae may also contribute to the CR
generation. The rate of occurrence, however, is 1/20 that of ccSN, if
integrated over the look back time to a higher redshift, and so their
contribution to extragalactic CRs would be insignificant.

We have other events that may contribute to CRs. 
The recent detection of a double neutron star merger (DNS) 
suggests a rate density of $\rho_{\rm DNS}\sim 1500~{\rm Gpc}^{-3}~{\rm yr}^{-1}$~\cite{TheLIGOScientific:2017qsa} with a large error. 
The kinetic energy of the ejecta is ${\mathcal E}_{\rm ej}\sim2\times10^{51}~{\rm erg}$. 
Their CR luminosity density, with the 10\% CR efficiency assumption, is 
\begin{equation}
Q_{\rm cr}^{\rm DNS}\sim3\times{10}^{44}~{\rm erg}~{\rm Mpc}^{-3}~{\rm yr}^{-1}~\epsilon_{\rm cr,-1}{\mathcal E}_{\rm DNS,51.3}, 
\end{equation}
which could give an order of magnitude of UHE CRs.

Gamma-ray bursts (GRB) are also suggested as sources of UHE CRs~\cite{Waxman:1995vg,Vietri:1995hs}. 
The local rate density is estimated to be around $\rho_{\rm
  GRB}\sim1~{\rm Gpc}^{-3}~{\rm yr}^{-1}$~\cite{Graham:2015rqk,
Wanderman:2009es}.
With the {\it Swift} sample, the isotropic-equivalent gamma-ray energy is ${\mathcal E}_{\rm iso\gamma}={10}^{52.3\pm0.7}$~erg~\cite{Wygoda:2015zua}, or
${\mathcal E}_{\rm iso\gamma}\sim{10}^{52.65}$~erg in the BATSE band. 
Then the gamma-ray luminosity of GRBs is  
\begin{equation}
Q_{{\rm MeV}\gamma}^{\rm GRB}\sim4\times{10}^{43}~{\rm erg}~{\rm Mpc}^{-3}~{\rm yr}^{-1}~{\mathcal E}_{{\rm iso}\gamma,52.65},
\end{equation}
which is again comparable to the luminosity density of UHE CRs above the ankle. 

It is considered that low-luminosity GRBs (LL GRB) form another
population of gamma-ray transients.  
The rate density of low-luminosity GRBs, if associated with transrelativistic 
supernovae with relativistic ejecta~\cite{Campana:2006qe,Soderberg:2006vh}, 
is $\rho_{\rm LLGRB}\sim100-1000~{\rm Gpc}^{-3}~{\rm yr}^{-1}$~\cite{Liang:2006ci,Sun:2015bda}. 
With an isotropic-equivalent radiation energy ${\mathcal E}_{{\rm iso}\gamma}\sim{10}^{50}$~erg gamma-ray
luminosity of low-luminosity GRBs is 
\begin{equation}
Q_{{\rm keV-MeV}\gamma}^{\rm LLGRB}\sim3\times{10}^{43}~{\rm erg}~{\rm Mpc}^{-3}~{\rm yr}^{-1}~{\mathcal E}_{{\rm iso}\gamma,50},
\end{equation}
which is comparable to that of canonical GRBs. 
Therefore, low-luminosity GRBs have often been considered as the CR sources at ultrahigh 
energies~\cite{Murase:2006mm,Zhang:2017moz}.

Starburst galaxies are a subclass of star-forming galaxies. In the IR
band, a significant fraction ($\sim5-100$\%) of star-forming galaxies
are claimed to be starbursts. 
In the local universe, however, only a few percent of the
star-formation occurs in starburst galaxies.  
The local CR luminosity density from star burst galaxies is 
\begin{equation}
Q_{\rm cr}^{\rm SBG}\sim1\times{10}^{45}~{\rm erg}~{\rm Mpc}^{-3}~{\rm yr}^{-1}.
\end{equation}
In starbursts CRs may efficiently interact with the ambient gas and lose almost all their energies via $pp$ interactions. 
Since $1/3$ of the CR energy goes to pionic gamma rays, the upper
limit on the gamma-ray luminosity density due to starburst galaxies is 
\begin{equation}
Q_{\rm GeV\gamma}^{\rm SBG}\sim3\times{10}^{44}~{\rm erg}~{\rm Mpc}^{-3}~{\rm yr}^{-1}, 
\end{equation}
which explains $\sim10-20$\% of the EGB if a spectral index of
$s=2.2$, as consistent with previous studies
(e.g.,~\cite{Ackermann:2012vca}).  Starbursts, if coexisting with AGN,
could enhance the contribution to the IGRB~\cite{Tamborra:2014xia}.

Stars close to the centre of galaxies may be tidally disrupted.
Some of such tidal disruption events (TDE) show relativistic jets with an
isotropic-equivalent x-ray luminosity
$L_{{\rm iso}\gamma}\sim10^{48}~{\rm erg}~{\rm s}^{-1}$. Based on the
Swift observation of Sw 1644+57, one may infer the rate density of
jetting TDEs to be
$\rho_{\rm TDE}\sim0.03~{\rm Gpc}^{-3}~{\rm yr}^{-1}$~\cite{2011Natur.476..421B}.  
We obtain their x-ray luminosity density
\begin{equation}
Q^{\rm TDE}_{{\rm keV-MeV}\gamma}\sim3\times{10}^{43}~{\rm erg}~{\rm Mpc}^{-3}~{\rm yr}^{-1}~{\mathcal E}_{{\rm iso}\gamma,54}, 
\end{equation}
where ${\mathcal E}_{{\rm iso}\gamma}=L_{{\rm iso}\gamma}\Delta T$ with $\Delta T\sim10^6$~s the duration of TDEs~\cite{2011Natur.476..421B}.
This number is again comparable to the UHE CR luminosity density, and hence TDEs can also be
considered as sources of UHE CRs~\cite{Farrar:2008ex,Farrar:2014yla}.

\subsection{Supermassive Black Holes}
\subsubsection{Active Galactic Nuclei}
Nonthermal x-ray emission arises from Comptonization of disc emission
in thermal coronae heated via magnetic reconnections.  X-ray
observations give the luminosity density of AGN in the $2-10$~keV
band~\cite{Ueda:2014tma},
\begin{equation}
Q^{\rm AGN}_{{\rm keV}\gamma}\simeq2.0\times{10}^{46}~{\rm erg}~{\rm Mpc}^{-3}~{\rm yr}^{-1}, 
\end{equation}
which is comparable to the CR luminosity density of star-forming
galaxies.  We note that a large part of the observed x-ray background is
ascribed to unresolved and obscured AGN.  

The bolometric luminosity of x-rays,
which is dominated by supermassive black holes with a standard
accretion disc or a radiatively more efficient slim disc, is
mostly of thermal origin, giving $Q^{\rm AGN}_{\rm bol}\simeq6.3\times{10}^{47}~{\rm erg}~{\rm Mpc}^{-3}~{\rm yr}^{-1}$~\cite{Ueda:2014tma}.
Recent observations indicate that quasars and Seyferts
ubiquitously possess fast outflows, which carry $\sim10$\% of the
bolometric luminosity~\cite{Tombesi:2012ad,Tombesi:2014waa}. 
Then, if $\epsilon_{\rm cr}\sim10$\% of the
kinetic luminosity is converted into CRs via the acceleration,
we may have 
\begin{equation}
Q^{\rm AGN}_{\rm cr}\sim6\times{10}^{45}~{\rm erg}~{\rm Mpc}^{-3}~{\rm yr}^{-1}~\epsilon_{\rm cr,-1}, 
\end{equation}
which is comparable to Eq.~(\ref{crbudget}) within an order of magnitude. 

AGN outflows driven by a central AGN
are also relevant for reacceleration in circumgalactic space: 
CRs may further be accelerated in powerful galactic winds.
As mentioned in the main text, the reacceleration in low-density regions 
could explain very high-energy CRs above the knee energy.

\subsubsection{Blazars}
The luminosity density of blazars is estimated using {\it Fermi}'s
gamma-ray data in the $0.1-100$~GeV range.  Blazars are divided
into two types, BL Lac objects and flat-spectrum radio quasars
(FSRQs).  The latter is also referred to as quasar-hosted blazars.
They are usually more luminous and more rapidly evolving than BL
Lacs. With a luminosity-dependent evolution model, the local gamma-ray
luminosity of BL Lac objects is~\cite{Ajello:2013lka}
\begin{equation}
Q^{\rm BL~Lac}_{{\rm GeV}\gamma}\simeq2.5\times{10}^{45}~{\rm erg}~{\rm Mpc}^{-3}~{\rm yr}^{-1}, 
\end{equation}
where $\sim50$\% errors in 1$\sigma$ is understood.  The gamma-ray
data show that BL Lacs have a weak evolution in redshift,
corresponding to $\xi_z\sim0.8$.  The gamma-ray energy density is
$\Omega_\gamma\sim{10}^{-10.0}$, which is comparable in order of
magnitude to that of the EGB.  It has been suggested that BL Lacs can
explain a significant fraction of the EGB.  A recent
analysis on the photon count distribution~\cite{TheFermi-LAT:2015ykq}
claimed that $86\pm14$\% of the EGB are explained by blazars.

Similarly, the local gamma-ray luminosity density of FSRQs is~\cite{Ajello:2013lka}
\begin{equation}
Q_{\rm FSRQ}^{{\rm GeV}\gamma}\sim2\times{10}^{44}~{\rm erg}~{\rm Mpc}^{-3}~{\rm yr}^{-1},
\end{equation}
with errors by a factor of 2. 
FSRQs show a strong redshift evolution: $\xi_z\sim8$.  
The gamma-ray energy density of FSRQs is estimated to be
$\Omega_\gamma\sim{10}^{-10.1}$, which may also account for a
significant fraction of the EGB.  At present epoch, the energy
densities of gamma rays from BL Lacs and FSRQs seem comparable.

\subsubsection{Misaligned Radio-Loud Active Galactic Nuclei}
The number density of misaligned radio-loud AGN~\cite{Willott:2000dh,vanVelzen:2012fn}
is $n_{\rm  RG}\sim10^{-5}-10^{-4}~{\rm Mpc}^{-3}$. 
Radio observations allow us to estimate the kinetic luminosity of jets 
$\sim{10}^{44}-{10}^{45}~{\rm erg}~{\rm s}^{-1}$~\cite{Willott:1999xa,Cavagnolo:2010jd},
using theoretically-motivated scaling relations.  
Assuming that $\epsilon_{\rm cr}\sim10$\% of the jet luminosity is converted to CRs, the CR luminosity density reads
\begin{equation}
Q^{\rm RG}_{\rm cr}\sim3\times{10}^{46}~{\rm erg}~{\rm Mpc}^{-3}~{\rm yr}^{-1}~\epsilon_{\rm cr,-1},
\end{equation}
which is comparable to or could even be larger than the CR luminosity density from star-forming galaxies. 
A large systematic uncertainty, however, comes from the used relationship between the radio and jet luminosities.

Radio galaxies are observed with {\it Fermi} in gamma rays. If we use
the empirical relation, e.g.,
$L_\gamma\propto L_{5~{\rm GHz}}^{1.16}$,
radio galaxies make a significant contribution to the EGB~\cite{Inoue:2011bm,Hooper:2016gjy}. 
It is found that $\Omega_{\gamma}\sim{10}^{-10.1}$, however, with a large uncertainty, corresponding to
\begin{equation}
Q_{\rm GeV\gamma}^{\rm RG}\sim4\times{10}^{44}~{\rm erg}~{\rm Mpc}^{-3}~{\rm yr}^{-1}. 
\end{equation}
It is likely that misaligned radio-loud AGN give a contribution to the
EGB comparable to that of blazars.  If we assume the beaming factor of
AGN jets $f_b\ll1$, the source number density of blazars is $f_b^{-1}$
times that of AGN, and their isotropic-equivalent
luminosity is enhanced by the same factor $f_b^{-1}$. Thus, $f_b$
cancels out.  Their gamma ray contribution being similar to that from
blazar is then naturally understood if the blazer and radio galaxies are in
fact the same entity.

\subsection{Hierarchical Structure Formation}
Extragalactic CRs may also be accelerated in clusters and groups of galaxies, 
where the intergalactic medium is heated
(e.g.,~\cite{Kang:1995xw,Murase:2008yt}). 
Accretion luminosity of  galaxy cluster  is  
$L\approx(\Omega_b/\Omega_m)GM\dot{M}/r_{\rm  vir}\simeq0.9\times{10}^{46}~{\rm erg}~{\rm s}^{-1}~M_{15}^{1/3}$,
where $M$ is the cluster (or group) mass and $r_{\rm vir}$ the virial radius.
The cluster number density above $10^{15}~M_\odot$ is $n_{\rm cl}\sim10^{-6}~{\rm Mpc}^{-3}$.
Assuming $\epsilon_{\rm cr}\sim10$\%, 
the CR luminosity density due to accretion shocks is on the order 
\begin{equation}
Q_{\rm cr}\sim3\times{10}^{46}~{\rm erg}~{\rm Mpc}^{-3}~{\rm yr}^{-1}~\epsilon_{\rm cr,-1},
\end{equation}
which is comparable to that of star-forming galaxies and radio galaxies within an order of magnitude.

As a result of hierarchical halo mergers, the kinetic energy is dissipated by shocks 
caused by galaxy and cluster mergers. The CR luminosity due to the halo mergers is~\cite{Yuan:2017dle},
\begin{equation}
Q_{\rm cr}\sim2\times{10}^{45}~{\rm erg}~{\rm Mpc}^{-3}~{\rm yr}^{-1}~\epsilon_{\rm cr,-1}, 
\end{equation}
which is comparable to CRs in starburst galaxies. 



\bibliography{kmurase.bib}



\end{document}